\documentclass [12pt]{article}
\usepackage{amsmath, amsthm, amssymb}
\usepackage{bm}
\usepackage{graphicx}
\usepackage{mathptmx}
\usepackage{fullpage}
\usepackage{authblk}
\usepackage{mathrsfs}
\usepackage{enumerate}
\usepackage{makecell}
\usepackage{bibentry, natbib}
\usepackage{float}
\usepackage[colorlinks=true, linkcolor=blue, citecolor=blue]{hyperref}
\newtheorem{Theorem}{Theorem}

\newtheorem{Remark}{Remark}

\newtheorem{Definition}{Definition}
\newtheorem{Proposition}{Proposition}
\newtheorem{Example}{Example}
\newtheorem*{Example*}{Example}
\usepackage{tikz}
\usepackage{pgflibraryarrows}
\usepackage{pgflibrarysnakes}
\usepackage{tablefootnote}
\usepackage{booktabs,caption,fixltx2e}
\usepackage[flushleft]{threeparttable}
\usepackage{multirow}
\usepackage{caption}
\emergencystretch=\maxdimen
\title{The Machiavellian frontier of stable mechanisms\thanks{This research is supported by the National Natural Science Foundation of China (Grant No. 72273043) and the College Students' Innovation and Entrepreneurship Training Program of East China University of Science and Technology (Grant No. S202310251190). A great debt is owed to Yajing Chen, Sambuddha Ghosh, Gaoji Hu, and Qianfeng Tang for detailed advice.
Special thanks to Sean Horen and the anonymous MSS journal referee for their insightful comments and patience with our previous submission.
All errors are our own responsibility.
}
}
\author{\small 
Qiufu Chen$^a$,
Yuanmei Li$^a$,
Xiaopeng Yin$^a$,
Luosai Zhang$^b$\footnote{Corresponding author:
Luosai Zhang.  Postal address: 777 Guoding Road, Shanghai, 200433, China. Email: zhangluosai@126.com},
Siyi Zhou$^a$
\\
\small $^a$School of Business, East China University of Science and Technology, Shanghai, China. 
\\
\small $^b$ School of Economics, Shanghai University of Finance and Economics, Shanghai, China\\
}

\begin{document}

\maketitle
%===============================================
\begin{abstract}

The impossibility theorem in \cite{roth1982economics} states that no stable mechanism satisfies strategy-proofness. 
This paper explores the Machiavellian frontier of stable mechanisms by weakening strategy-proofness.  
For a fixed mechanism $\varphi$ and a true preference profile $\succ$, a $(\varphi,\succ)$-boost mispresentation of agent $i$ is a preference of $i$ that is obtained by (i) raising the ranking of the truthtelling assignment ${\varphi}_i (\succ)$, and (ii) keeping rankings unchanged above the new position of this truthtelling assignment. 
We require a matching mechanism $\varphi$ neither punish nor reward any such misrepresentation, and define such axiom as $\varphi$-boost-invariance.\footnote{The axiom of $\varphi$-boost-invariance \`a la \cite{chen2024machiavellian}( named as `truncation-invariance' there)  is constructed based on the `$(\varphi,\succ)$-boost misrepresentation'  \`a la Chen (2017) ( named as `truncation strategy' there).} 
This is strictly weaker than requiring strategy-proofness.
We show that no stable mechanism $\varphi^{*}$ satisfies $\varphi^{*}$-boost-invariance.
Our negative result strengthens the Roth Impossibility Theorem.

%We require that no agent can obtain a different assignment by any such misrepresentation. 
%This is strictly weaker than requiring strategy-proofness.
%Our main result strengthens the Roth Impossibility Theorem to the following: No stable mechanism is invariant to such misrepresentation.

%We show that every stable mechanism is susceptible to such misreprentation, i.e. for every stable mechanism $\varphi$ some agent $i$ can obtain a different assignment at some $\succ$ by such a $(\varphi,\succ)$-misrepresentation. The negative result strengthens the Roth Impossibility Theorem.

\vspace{3mm}
\vspace{3mm}


\end{abstract}

\newpage
%===============================================
\section{Introduction}
%===============================================

This paper investigates the one-to-one matching problem without monetary transfers in the marriage market introduced by \citet{gale1962college}. 
This market consists of a set of men and a set of women, and each agent holds a strict preference for being matched either with another agent from the opposite set or with oneself.
We consider the principle of ``stability" to determine a successful matching. 
Stability means there are no pair of agents who would prefer to be matched with each other over their current partners (known as ``blocking pairs"), nor is there an individual who prefers being matched with oneself over the assigned partner.

\citet{roth1982economics} and \citet{roth1984misrepresentation} extend the model by introducing
the preference revelation game where agents present their preferences to a matchmaker and the matchmaker apply a mechanism to select a matching according to their presentations.
A stable matching mechanism always selects a stable matching with respect to the stated preferences. 
In a strategy-proof matching mechanism, every agent prefers to reveal truthfully no matter how others reveal their preferences.\footnote{Note that in the preference revelation game, truthful revelation constitutes a Nash equilibrium only if it is also a dominant strategy equilibrium \citep{maskin2002implementation}. Thus, in a strategy-proof mechanism, truthful revelation is a Nash equilibrium if and only if it is a dominant strategy equilibrium.}
Thus, an ideal matching mechanism should be both stable and strategy-proof, ensuring that the matching it selects is stable with respect to both the stated and true preferences.
However,
\citet{roth1982economics} offers the Roth Impossibility Theorem: 
There is no stable mechanism that satisfies stategy-proofness.

%Some studies investigate the constraints on preference domains necessary for ensuring strategy-proofness in stable mechanisms. Notable examples include the ``top dominance" condition \citep{alcalde1994top} and the ``no-detour" condition \citep{akahoshi2014necessary}.
%A stable mechanism is strategy-proof if and only if there exists a singleton core in the market.\footnote{This result is from Theorem 3.19 in \cite{roth1992two} and corollary 4 in \cite{Ma1995}.} 
%Therefore, this branch of literature is also related to the literature on identifying necessary and sufficient conditions for the singleton core, such as \citet{eeckhout2000equilibria}, \citet{clark2006efficiency} and \citet{akahoshi2014singleton}.

Strategy-proofness is a particularly strong requirement for stable mechanisms. 
Firstly, in order to make the Nash equilibrium outcome stable with respect to true preferences, truthful revelation is not a necessary requirement. \citet{roth1984misrepresentation} shows that in the man-optimal stable mechanism or the woman-optimal stable mechanism there exists a misrepresentation (Nash) equilibrium that is stable with respect to the true preferences. 
Secondly, the strategy-proofness in Roth Impossibility Theorem implicitly excludes manipulation on strategy domain such as going for outside options \citep{Sirguiado2024}. \cite{Sirguiado2024} demonstrate that when agents have no outside options, and this is known to the matchmaker, \cite{roth1982economics}'s impossibility theorem applies if and only if there are at least three agents on each side.
Lastly, empirical evidence weakly supports strategy-proofness. 
\citet{charness2009origin} and \citet{esponda2014hypothetical} demonstrate that individuals struggle with hypothetical reasoning, even in single-agent decision problems. 
Therefore, relaxation of strategy-proofness has emerged in literature, such as the concept of obvious manipulation introduced by \citet{troyan2020obvious}.
 
Previous studies show that some stable mechanisms can satisfy certain desirable properties that are weaker than strategy-proofness.
Note that strategy-proofness is achieved through implementation via a dominant strategy equilibrium. 
Some research weakens strategy-proofness by relaxing the requirement of dominant strategy equilibrium.
For example, \citet{dubins1981machiavelli} show that any ``strong Nash equilibrium" outcome of the man-optimal stable mechanism is woman-optimal stable. Similarly, \citet{Ma1995} shows that any ``rematching-proof equilibrium" outcome of the man-optimal stable mechanism is woman-optimal stable, and \citet{Alcalde1996} shows that any ``dominance solvable equilibrium" outcome of the man-optimal stable mechanism is woman-optimal stable. 
All these implementation properties are introduced based on refinements of the Nash equilibium involve misrepresentations.
In contrast, our axiom of ``$\varphi$-boost-invariance'', which is also a weaker property than strategy-proofness (See Proposition \ref{prop:1}), is not a property on implementation.
% with some notion of equilibrium.
Nevertheless, their result overturn the Roth Impossibility Theorem, while our result strengthens the Roth Impossibility Theorem.

%In a matching mechanism $\varphi$, an agent's {\it $\varphi$-assignment under truth} ($\varphi$-AT) is defined as his/her assignment by $\varphi$ when all agents reveal their preferences truthfully.
%% A strategy maps an agent's true preference to a stated preference. 
For a fixed mechanism $\varphi$, an agent $i$'s {\it $(\varphi,\succ)$-boost misrepresentation} is a preference of $i$ obtained by raising the ranking of $\varphi_i(\succ)$---agent $i$'s assignment under $\varphi$ when everyone reveal truthfully, and keeping rankings unchanged above the new position of the truthtelling assignment. 
A mechanism $\varphi$ is {\it $\varphi$-boost-invariant} if no agent can obtain a different assignment by any such misrepresentation.
This axiom is first introduced by \cite{chen2024machiavellian} to characterize the top trading cycles (TTC) mechanism in the housing market problem \`a la \citep{shapley1974cores}.\footnote{$\varphi$-boost-invariance is referred to as ``truncation-invariance" in \citet{chen2024machiavellian}. The new name distinguishes our $(\varphi, \succ)$-boost misrepresentation from the truncation strategy in \citet{roth1991incentives}.} 
Note that the $\varphi$-boost-invariance is a slightly weaker requirement than \cite{takamiya2001coalition}'s individual monotonicity and can be viewed as an individual version of \cite{chen2017new}'s rank monotonicity. However, our motivations differ. Individual monotonicity and rank monotonicity are developed by weakening Maskin-monotonicity \citep{maskin1999nash} to characterize the TTC mechanism and the deferred acceptance (DA) mechanism respectively. 
Maskin-monotonicity is motivated by the idea that a desirable social choice rule \( f \) should satisfy the following: if for any two preference profiles \( R \) and \( R' \), and for any alternative \( x \) such that \( f(R') = x \), if for all individuals \( i \), \( x \) is ranked the same or higher in \( R_i \) than in \( R_i' \), then \( f(R) \) must also be \( x \). 
In contrast, $\varphi$-boost-invariance is motivated by the principle that a matching mechanism should neither punish nor reward any $\varphi$-boost misrepresentation.
%\footnote{Intuitively, cooperation with the matchmaker should be rewarded, while misrepresentations should not.}

This paper is also closely related to the literature on restricting the strategy domains for stable mechanisms, such as the ``protective strategy" by \citet{barbera1995protective} and the ``truncation strategy" by \citet{roth1991incentives}. 
\citet{barbera1995protective} introduce the protective strategy (lexical maximum strategy) to describe situations where agents exhibit extreme risk aversion through binary comparisons between strategies. 
They show that the man-optimal stable mechanism and the woman-optimal stable mechanism are both strategy-proof under the restriction of the strategy domain to protective strategies. 
\citet{roth1991incentives} consider truncation strategy, where an agent raises the ranking of oneself relative to their true preference while keeping rankings unchanged above the new position of oneself. They show that any matching achievable through an arbitrary misrepresentation by an agent can also be achieved through a truncation strategy of that agent.\footnote{Following \citet{roth1991incentives}, the truncation strategy has been theoretically studied by \cite{mongell1991sorority}, \citet{roth1999truncation}, \citet{Jaramillo2013}, \citet{castillo2016truncation}, and \citet{coles2014optimal}, and experimentally by \citet{coles2014optimal} and \citet{castillo2016truncation}.} While these papers restrict the strategy domains to investigate the strategy-proofness of stable mechanisms, this paper's axiom of $\varphi$-boost-\textit{invariance} specifies how a matching mechanism should respond to a $(\varphi, \succ)$-boost misrepresentation.

This paper is structured as follows: We introduce the model in section \ref{sec:2}, present our main theorem in section \ref{sec:3} and conclude in section \ref{sec:5}.

%===============================================
\section{Model}\label{sec:2}
%===============================================

Our model is that of \cite{gale1962college}. 
A {\bf marriage market} is a tuple $(M, W, \succ)$.
Agents are categorized into two disjoint and finite sets: men ($M$) and women ($W$);
members of these sets are denoted by the corresponding lowercase letters. 
The set of all agents is denoted by $\mathcal{I}= M \cup W$. 
For each $i\in \mathcal{I}$, let $X_i$ be the {\bf opposite set} of $i$; that is, $X_i=M$ if $i\in W$, and $X_i=W$ if $i\in M$.
Each agent can match with an agent from $X_i\cup\{i\}$.

We assume that each man $m$ has a strict \textbf{preference} $\succ_{m}$ over $W\cup\{m\}$, and each woman $w$ has a strict preference $\succ_{w}$ over $M\cup\{w\}$. 
Define $\succsim_i$ as the {\bf weak preference} induced by $\succ_i$, where $j\succsim_i k$ if and only if either $j\succ_i k$ or $j=k$.
An agent $j$ from one side is \textbf{acceptable} to another agent $i$ from the other side if $i$ prefers to be matched with $j$ rather than with himself/herself, i.e., $j$ is acceptable for $i$ if and only if $j\succsim_i i$.
Conversely, an agent $j$ from one side is \textbf{unacceptable} to another agent $i$ from the other side if $j$ is not acceptable to $i$.
The {\bf preference profile} is $\succ := (\succ_{i})_{i\in\mathcal{I}}$. 
The set of all possible preferences of agent $i$ is $\mathcal{P}_i$ and the set of all possible preference profiles is $\mathcal{P} := (\mathcal{P}_i)_{i\in \mathcal{I}}$. 
A \textbf{matching} is a function that selects a partner (can be oneself) for each agent.
Formally, a matching $\mu: M\cup W\rightarrow M\cup W$ should satisfy:
(i) $\mu({\mu(i)})=i$ for each $i\in M\cup W$;
(ii) $\mu(m)\in W\cup\{m\}$; $\mu(w)\in M\cup\{w\}$.
Restriction (i) ensures bilateral matching, while (ii) indicates that an agent can either match with someone from the opposite set or with oneself.
We denote the set of all possible matchings by $\mathcal{M}$. Henceforth we use $\mu_i$ instead of $\mu(i)$.

A pair $(m,w)\in M \times W$ is said to {\bf block} a matching $\mu$ if $m\succ_{w}\mu_w$ and $w\succ_{m}\mu_m$.
An agent $i\in M\cup W$ is said to {\bf block} a matching $\mu$ if $i \succ_{i}\mu_i$.
A matching $\mu$ is said to be {\bf individually rational} if it cannot be blocked by any agent.
A matching $\mu$ is said to be {\bf stable} if it cannot be blocked by any agent or pair.
We denote the set of all stable matchings as $\mathcal{M}^{S}$.
A stable matching $\mu\in \mathcal{M}^{S}$ is {\bf $M$-optimal} if every man prefers it at least as much as any other stable matching. Formally,  
%a stable matching $\mu$ is $M$-optimal if 
for each $\mu'\in\mathcal{M}^{S}$ and each $m\in M$, we have $\mu_m \succsim_m \mu'_m$. 
A stable matching $\mu\in \mathcal{M}^{S}$ is {\bf $W$-optimal} if every woman prefers it at least as much as any other stable matching. 
Formally,  
%a stable matching $\mu$ is $W$-optimal if 
for each $\mu'\in \mathcal{M}^{S}$ and each $w\in W$, we have $\mu_w \succsim_w \mu'_w$.\footnote{For the existence and uniqueness of the $M$-optimal stable matching and $W$-optimal stable matching, see \cite{gale1962college}. They develop the man-proposing (woman-proposing) deferred acceptance (DA) algorithm to find the unique $M$-optimal ($W$-optimal) stable matching for any market.}

A \textbf{matching mechanism}, employed by the matchmaker, prescribes a matching based on the stated preferences from all agents.
Formally, a mechanism $\varphi$ is an outcome function $\varphi: \mathcal{P} \rightarrow \mathcal{M}$, which finds a matching for each stated preference profile.  
%Let $\psi$ denote the set of all possible mechanisms. 
Thus, a {\bf strategy} of an agent $i\in\mathcal{I}$ in a matching mechanism (or simply a `mechanism') is a function $\sigma_i: \mathcal{P}_i \rightarrow \mathcal{P}_i$,
and a {\bf strategy profile} is thus a function $\sigma: \mathcal{P} \rightarrow \mathcal{P}$.\footnote{Note that in order to distinguish the true preferences from the {stated preferences} by agents within a mechanism, we use the label ``$\succ$'' for each true preference and the label ``$P$'' for each stated preference.}
Let $R_i$ denote the weak preference induced  by $P_i$. 
%where $j\ R_i\ k$ if and only if either $j\ P_i\ k$ or $j=k$.
A {\bf truthful revelation strategy} $\sigma^{*}_i$ of agent $i$ is a strategy such that $\sigma^{*}_i(\succ_i) = \succ_i$ for each $\succ_i \in \mathcal{P}_i$.

Let $\varphi(P)$ denote the matching selected by $\varphi$ for the stated preference profile $P$, and let $\varphi_i(P)$ denote the partner of agent $i$ at the matching $\varphi(P)$. 
A mechanism $\varphi$ is {\bf individually rational} if the matching it selects, i.e., $\varphi(P)$, is individually rational with respect to agents' stated preferences. 
A mechanism $\varphi$ is {\bf stable} if the matching $\varphi(P)$ it chooses is stable with respect to agents' stated preferences (i.e., $\varphi(P)$ is stable when the true preference profile is $P$).
%We denote the set of all possible stable mechanisms as $\psi^{S}$.
A stable mechanism  is {\bf $M$-optimal} (resp, {\bf $W$-optimal}) if the matching $\varphi(P)$ it chooses is {$M$-optimal} (resp, {$W$-optimal}) with respect to the stated preferences. 
We denote the $M$-optimal stable mechanism as $\varphi^{M}$ and the $W$-optimal stable mechanism as $\varphi^{W}$.
%We say a mechanism is a \textbf{DA-derived mechanism} if it is outcome-equivalent to either the $M$-optimal stable mechanism or the $W$-optimal stable mechanism for each stated preference profile.
%Formally, $\varphi^*$ is a \textbf{DA-derived mechanism} if 
%\[\varphi^* \in \{\varphi \in \psi | \varphi(P) \in \{\varphi^M(P), \varphi^W(P)\}, \forall P \in \mathcal{P}\}.\]

A mechanism $\varphi$ is \textbf{strategy-proof} if the truthful revelation strategy is a dominant strategy for each agent in the mechanism. 
Formally, $\varphi$ is strategy-proof if for each profile $\succ\in\mathcal{P}$, each agent $i\in \mathcal{I}$, and $P'=(P'_i,P'_{-i})\in\mathcal{P}$, we have $\varphi(\succ_i,P'_{-i})\ \succsim_i\ \varphi(P'_{i}, P'_{-i})$.\footnote{A mechanism $\varphi$ is \textit{Nash implementable with truthful revelation} if for each $\succ\in\mathcal{P}$, $i\in \mathcal{I}$, and $P'_i$ is an arbitrary stated preference from agent $i$, we have $\varphi_i(\succ_i, \succ_{-i})\ \succ_i\ \varphi_i(P'_{i},\succ_{-i})$.
Note that Nash implementable with truthful revelation is equivalent to strategy-proofness for revealed preference mechanisms (\cite{maskin2002implementation}).}

For each stated preference $P_i \in \mathcal{P}_i$ of an agent $i$, with some abuse of notation,  we denote by $P_i(j)$ the \textbf{rank} of $i$'s potential partner $j\in X_i$ at $P_i$, where for $k\in\{1,2,\cdots,|X_i|+1\}$, $P_i(j)=k$ means that $j$ is the $k^{th}$ preferred assignment in agent $i$'s stated preference $P_i$.
%We refer to $\varphi_i(\succ)$, i.e., $i$'s assignment selected by the given mechanism $\varphi$ when all agents reveal truthfully,  as agent $i$'s {\bf $(\varphi,\succ)$-assignment}.

\begin{Definition}[$(\varphi,\succ)$-boost misrepresentation]
For a fixed mechanism $\varphi$, and a fixed preference profile $\succ$, a {\bf $(\varphi,\succ)$-boost misrepresentation} of agent $i$ is a preference $P_i$ such that: 
\[
j \, P_i \, \varphi_i(\succ) \Rightarrow P_i(j)= \; \succ_i(j)\mbox{,  }\forall j \in X_i\cup \{i\},
\]
where $X_i$, the opposite set of $i$, is defined by $X_i=M$ if $i\in W$, and $X_i=W$ if $i\in M$.
\end{Definition}

Note that the definition above implies that in $P_i$: The rank of $\varphi_i(\succ)$ in $P_i$ is not lower than that in $i$'s true preference $\succ_i$; Options below $\varphi_i(\succ)$ in $P_i$ could have any ranking. 

\begin{Remark}
An agent's $(\varphi, \succ)$-boost misrepresentation is not a strategy of this agent, because it depends not only on the true preference (type) of the agent but also on the true preference profile $\succ$ and his/her assignment under $\varphi$ when everyone reveals truthfully.
\end{Remark}

Now we are ready to introduce our axiom.

\begin{Definition}[$\varphi$-boost-invariance]
A mechanism $\varphi$ satisfies \textbf{$\varphi$-boost-invariance} if for each profile $\succ\in\mathcal{P}$, each agent $i\in \mathcal{I}$, and each $(\varphi, \succ)$-boost misrepresentation $P'_i$ from $i$, we have 
$$\varphi_i(\succ)= \varphi_i(P'_i, \succ_{-i}).$$
\end{Definition}

We shall show later that $\varphi$-boost-invariance is a weaker axiom than strategy-proofness.

\subsection{Truncation Strategy vs. $(\varphi, \succ)$-Boost Misrepresentation}\label{sec:2.1}

%{\color{purple}
%This paper is also related to the literature on trucation strategies.
Previous literature identifies two types of misrepresentations both named ``truncation strategy":

\begin{itemize}
\item A {\it truncation strategy} of an agent $i$ introduced by \cite{roth1991incentives} is defined by:
(i) raising the ranking of the agent oneself relative to the true preference, and
(ii) keeping the rankings unchanged above the new position of the agent oneself.
\item A {\it truncation strategy} of an agent $i$ introduced by \citet{chen2017new}, which is renamed as {\it $(\varphi,\succ)$-boost misrepresentation} is a preference of $i$ that is obtained by:
(i) raising the ranking of the assignment that would result under the given mechanism $\varphi$ if everyone announced truthfully (relative to the true preference), and
(ii) keeping rankings unchanged above the new position of this truthtelling assignment.
\end{itemize}

The following example demonstrates that the restriction of misrepresentations to truncation strategies and the restriction to $(\varphi,\succ)$-boost misrepresentations do not imply each other: for a fixed mechanism with a fixed true preference profile,  a truncation strategy may not be a $(\varphi,\succ)$-boost misrepresentation, and vice versa.
%To better understand the differences between $\varphi$-AT floating strategy and truncation strategy, let us consider the following example.
\begin{Example}

Consider a marriage market with two men, $M=\{m_1, m_2\}$, and two women, $W=\{w_1, w_2\}$. The true preference profile, $\succ$, is listed (on the left-hand side) as follows.

\begin{center}
\begin{tabular}{cc|cccccc|cc}

$P_{m_1}$ & $P_{m_2}$  & $P_{w_1}$ & $P_{w_2}$ && $P^{1}_{m_1}$ & $P^{2}_{m_1}$ & $P^{3}_{m_1}$ & $P^{a}_{m_1}$ & $P^{b}_{m_1}$ \\
\cline{1-4}\cline{6-10}
$w_2$         & $w_1$   & $m_1*$         & $m_2*$  & & $w_2$  & $\boxed{m_1}$  & $\boxed{m_1}$    & $\boxed{w_1}$  & $\boxed{w_1}$   \\

$w_1*$         & $w_2*$   & ${m_2}$    & ${m_1}$  && $\boxed{m_1}$          & $w_1$ & $w_2$ & $w_2$ & $m_1$\\

$m_1$        & $m_2$       & $w_1$      & $w_2$   && ${w_1}$         & $w_2$ & $w_1$ & $m_1$ & $w_2$ \\

\end{tabular}
\end{center}

The $W$-optimal stable mechanism  $\varphi^{W}$ selects the matching labeled with ``$*$''  above.
Thus, as listed on the right-hand side above, agent $m_1$'s truncation stategies here are: $P^{1}_{m_1}$ , $P^{2}_{m_1}$, $P^{3}_{m_1}$, but $m_1$'s $(\varphi^{W},\succ)$-boost misrepresentations are: $P^{a}_{m_1}$, $P^{b}_{m_1}$.
\end{Example}

These two types of manipulations have been explored based on different motivations.
Any matching that can be achieved through an arbitrary misrepresentation by an agent can also be obtained through a truncation strategy of that agent \citep{roth1999truncation}. Building on this property, \citet{Jaramillo2013} investigates truncation and dropping strategies in many-to-many matching mechanisms.\footnote{A dropping strategy is a stated preference obtained by removing acceptable partners without reshuffling. Each truncation strategy is also a dropping strategy.
Applying \citet{Jaramillo2013}'s results to one-to-one markets suggests that if agents understand the exhaustiveness of truncation or dropping correspondences, these agents will reveal truthfully about the relative rank order of the listed partners.} 
\citet{chen2017new} designed the $(\varphi,\succ)$-boost misrepresentation to introduce the axiom of rank monotonicity to characterize the DA mechanism. Based on the concept of $(\varphi, \succ)$-boost misrepresentation, \citet{chen2024machiavellian} define the axiom of $\varphi$-boost-invariance to characterize the TTC mechanism in the housing market problem of \cite{shapley1974cores}.
%These two types of manipulations on the strategy domains have been explored in the literature based on different motivations.
%\begin{itemize}
%\item Note that any matching that can be obtained through an arbitrary misrepresentation from an agent can also be achieved through a truncation strategy of this agent \citep{roth1999truncation}.
%%\cite{mongell1991sorority},\citet{roth1991incentives}, \citet{roth1999truncation}, 
%Based on the good property above, \citet{Jaramillo2013} investages the truncation strategies and dropping strategies in the many-to-many matching mechanisms.\footnote{A dropping strategy is a list
%that is obtained from an agent’s true preference list by removing acceptable partners i.e., no reshuffling. Obviously, each truncation strategy is also a dropping strategy.}
%Note that by applying \citet{Jaramillo2013}'s result on many-to-many matching problem to one-to-one markets, 

We adopt the $(\varphi,\succ)$-boost misrepresentation instead of the truncation strategy to construct the axiom of $\varphi$-boost-invariance for two reasons:

\begin{enumerate}
\item \textbf{Theoretical Exploration}: We need an axiom that is strictly weaker than strategy-proofness to explore the Machiavellian frontier of stable mechanisms, where \cite{roth1982economics}'s impossibility theorem still holds. If we construct an axiom of \textit{truncation-invariance} based on the truncation strategy \`a la \cite{roth1991incentives},\footnote{That is, a mechanism is truncation-invariant if it still assigns an agent the partner where every agent reveals truthfully when this agent unilaterally adopts a truncation strategy \`a la \cite{roth1991incentives}.} then by definition it is a strictly stronger axiom than strategy-proofness with the restricted domain of truncation strategies \`a la \cite{roth1991incentives}. Note that strategy-proofness under such a restricted domain is equal to strategy-proofness, since \cite{roth1999truncation} shows any matching achieved through an arbitrary misrepresentation by an agent can also be obtained through a truncation strategy of that agent. Therefore, truncation-invariance is not a weaker axiom than strategy-proofness, while as we will show in Proposition \ref{prop:1}, $\varphi$-boost-invariance is a strictly weaker axiom than strategy-proofness.

\item \textbf{Experimental Intuition}: We hope to find an axiom to intuitively explain phenomena observed in experimental studies. \cite{castillo2016truncation} shows that when an agent is limited to truncation strategies in an experimental setting, the agent focuses more on the rank of the best achievable match rather than acting optimally.\footnote{They find that the lower the best achievable match is in the agent's preference, the higher the probability the agent will truncate the list optimally.}
\end{enumerate}

%===============================================
\section{The Main Result}\label{sec:3}
%===============================================
\cite{roth1982economics} proposes a famous impossibility theorem as follows:

\vspace{2mm}

\noindent{\bf Theorem 0} (\cite{roth1982economics}, Theorem 3).\label{lem:1}
No stable mechanism satisfies strategy-proofness. 

\vspace{2mm}

In this section, we relax the notion of strategy-proofness, aiming to explore the Machiavellian frontier of stable mechanisms.
Proposition 1 in \cite{chen2024machiavellian} shows that in the housing market problem  of \cite{shapley1974cores}, the strategy-proofness of an allocation rule implies $\varphi$-boost-invariance of that rule.
The following proposition shows that the same holds for one-to-one matching mechanisms.
Note that the logic of this proof follows that in \cite{chen2024machiavellian}. It is offered because our model setting is different: \cite{chen2024machiavellian} applies to the housing market model, while ours is a marriage market.

\begin{Proposition}\label{prop:1}
If a mechanism is strategy-proof, then it is $\varphi$-boost-invariant.
\end{Proposition}

\begin{proof}
Suppose $\varphi$ satisfies strategy-proofness, but it violates $\varphi$-boost-invariance. 

Then, there exists $P\in\mathcal{P}$, $i\in\mathcal{I}$, and $P'_i$ is a $(\varphi,\succ)$-boost misrepresentation of agent $i$, such that $\varphi_i(\succ) \neq \varphi_i(P'_i, \succ_{-i})$.

By strategy-proofness of $\varphi$, we have $\varphi_i(\succ) \succ_i \varphi_i(P'_i, \succ_{-i})$.
 
Since $P'_i$ is a $(\varphi,\succ)$-boost misrepresentation of $i$, we have $\varphi_i(\succ) P'_i \varphi_i(P'_i, \succ_{-i})$, which violates the strategy-proofness of $\varphi$. 
Note that in the above argument, we consider $(P'_i, \succ_{-i})$ as the true preference profile, and $\succ$ as the “manipulated” preference profile.
\end{proof}

The immediate acceptance (IA) mechanism introduced by \cite{abdulkadirouglu2003school} satisfies $IA$-boost-invariance but violates strategy-proofness; so the former is strictly weaker than the latter.

\vspace{2mm}

\noindent{\bf The Immediate Acceptance (IA) mechanism.}
The IA mechanism finds a matching $IA(P)$ for each profile $P$ through the following man-proposing (woman-proposing) IA algorithm:

\noindent\textbf{\emph{Step $1$.}} Each man (woman) proposes to his (her) favorite woman (man). Each woman (man) then permanently accepts the proposal from her favorite partner and rejects the other proposals. The men (women) who are accepted by some women (men) are removed with their partners.

\noindent\textbf{\emph{Step $k$ ($k\geq 2$).}} Each remaining man (woman) proposes to his $k^{th}$ preferred woman (man). Each remaining woman (man) then permanently accepts the proposal from her (his) favorite partner and rejects the other proposers.

\noindent The algorithm terminates in a finite number of steps when all agents have been removed.

From the procedure of the IA algorithm, it is evident that the IA rule satisfies $IA$-boost-invariance. The IA rule is exactly the so-called Boston mechanism. It is well-known that such a mechanism is not strategy-proof (\cite{abdulkadirouglu2003school}).

\subsection{Impossibility Theorem with $\varphi$-Boost-Invariance}

By weakening strategy-proofness into $\varphi$-boost-invariance, we present an impossibility theorem asserting the nonexistence of an $\varphi$-boost-invariant stable mechanism, thereby extending the Roth Impossibility Theorem. 

%======================================================================
\begin{Theorem}\label{thm:impossibility}
No stable mechanism satisfies $\varphi$-boost-invariance.
\end{Theorem}
%======================================================================
\begin{proof}Fix a stable and $\varphi$-boost-invariant mechanism $\varphi$, if any. We show that $\varphi$ is the same as the woman-optimal stable mechanism $\varphi^{W}$.
Next, we show that $\varphi^{W}$ is not $\varphi$-boost-invariant -- a contradiction.

{\bf Step 1. }  We show that $\varphi$ is the same as $\varphi^{W}$.

Suppose there exists a profile $\succ\in\mathcal{P}$ such that $\varphi(\succ)\neq\varphi^{W}(\succ)$.
Then, there must exist a woman $w$ such that $\varphi^{W}_w(\succ)\succ_w \varphi_{w}(\succ)\succsim_w w$.
Consider two preferences $P'_w, P''_w\in\mathcal{P}_{w}$ of $w$ as follows:
\begin{itemize}
\item $P'_w$ satisfies:

$(\romannumeral 1)\mbox{ } P'_w(m)=\ \succ_w (m), \mbox{ if } m \succsim_w \varphi^{W}_w(\succ)$;

$(\romannumeral 2)\mbox{ } P'_w(w)=P'_w (\varphi^{W}_w(\succ))+1$.

\item $P''_w$ satisfies:

$(\romannumeral 1)\mbox{ } P''_w(m)=\ \succ_w (m), \mbox{ if } m \succsim_w \varphi_w(\succ)$;

$(\romannumeral 2)\mbox{ } P''_w(w)=P''_w (\varphi_w(\succ))+1, \mbox{ if } \varphi_w(\succ) \neq w$.

\end{itemize}

To be specifically $\succ_w$, $P'_w$ and $P''_w$ can be listed as in the following table:
\begin{center}
\begin{tabular}{c|c|c}

$\succ_{w}$            & $P'_{w}$                & $P''_{w}$  \\
\hline
$\vdots$               & $\vdots$                & $\vdots$   \\

$\varphi^{W}_w(\succ)$ & $\varphi^{W}_w(\succ)$  & $\varphi^{W}_w(\succ)$ \\

$\vdots$               & $w$                     & $\vdots$    \\
$\varphi_w(\succ)$     & $\vdots$               & $\varphi_w(\succ)$\\
$\vdots$               &                 & $w$   \\
$w$                    &                & $\vdots$ \\
$\vdots$               &                & \\
\end{tabular}
\end{center}

%We denote $P'=(P'_w, \succ_{-w})$ and $P''=(P''_w, \succ_{-w})$.
By the construction of \(P'_w\),\footnote{Note that $P'_w$ is a truncation strategy of $\succ_w$ \`a la \citet{roth1991incentives}.}  it can be inferred that the woman-proposing deferred acceptance (WDA) algorithm (\cite{gale1962college}) will select the same matching when the stated preference profile is \((P'_w, \succ_{-w})\) as it does with \(\succ\). 
Theorem 2 in \cite{roth1982economics} shows that
the unique $W$-optimal stable matching can be find through the procedure of
the WDA algorithm (\cite{gale1962college}).
Therefore, we have 
$$\varphi^{W}(P'_w, \succ_{-w})=\varphi^{W}(\succ).$$

Thus, by the individual rationality and stability of $\varphi^M$ and the fact that $\varphi^{W}$ selects the best stable matching for each woman (Theorem 2 in \cite{roth1982economics}) , we have :
\[
\varphi^{M}_w(P'_w, \succ_{-w})\in \{w, \varphi^{W}(P'_w, \succ_{-w})\}.
\]
Note that $\varphi^{M}_w(P'_w, \succ_{-w})\neq w$ otherwise the relation above will violates the rural hospital theorem in \cite{roth1986allocation}. 
Therefore, we have $\varphi^{W}_w(P'_w, \succ_{-w})=\varphi^{M}_w(P'_w, \succ_{-w})=\varphi^{W}_w(\succ)$.
Thus, by the lattice theorem (Theorem 2.16 in \cite{roth1992two}, page: 36-39), we have 
$$\varphi_w(P'_w, \succ_{-w})=\varphi^{W}_w(\succ).$$

Then, we can derive $\varphi_w(P''_w, \succ_{-w})$ in two approaches as follows:

{\bf Approach 1. } By construction, $P''_w$ is a $(\varphi,\succ)$-boost misrepresentation of agent $w$. Therefore, the $\varphi$-boost-invariance of $\varphi$ implies 
$$\varphi_w(P''_w, \succ_{-w})=\varphi_w(\succ).$$

{\bf Approach 2. } By construction, $P''_w$ is a $(\varphi, (P'_w, \succ_{-w}))$-boost misrepresentation of agent $w$.
Therefore, the $\varphi$-boost-invariance of $\varphi$ implies 
$$\varphi_w(P''_w, \succ_{-w})=\varphi_w(P'_w, \succ_{-w})=\varphi^{W}_w(\succ).$$

Approach 1 contradicts with approach 2 since $\varphi_w(\succ)\neq\varphi^{W}_w(\succ)$.

\vspace{4mm}

{\bf Step 2. }We show through the following example that $\varphi^{W}$ is not $\varphi^{W}$-boost-invariant.
%===============================================

\vspace{2mm}

%%%%%%%%%%%%%%%%%%%%%%%%%%%%%%%%%%%%%%%%%%%%%%%%%%%%%%%%%%%%%%%%%%%%%%%
%%  Example 1
%%%%%%%%%%%%%%%%%%%%%%%%%%%%%%%%%%%%%%%%%%%%%%%%%%%%%%%%%%%%%%%%%%%%%%%

Let $M= \{m_1, m_2, m_3\}$, $W=\{w_1, w_2, w_3\}$. 
The true preferences are listed on the left-hand side below. A $(\varphi,\succ)$-misrepresentation of $m_2$ is listed on the right-hand side below.

\begin{center}
\begin{tabular}{ccc|ccccccc}

$\succ_{m_1}$ & $\succ_{m_2}$ & $\succ_{m_3}$ & $\succ_{w_1}$ & $\succ_{w_2}$ & $\succ_{w_3}$  &&& $P_{m_2}^{'}$ &\\
\cline{1-6}\cline{9-9}
$w_2$           & $\boxed{w_1}$  & $w_1$           & $m_1*$         & $m_2$         &$m_2*$ &&& $w_1$ &\\

$\boxed{w_3}$   & $w_2$          & $w_3$           & $\boxed{m_2}$ & $\boxed{m_3}*$ &$\boxed{m_1}$ &&& $w_3*$ &\\

$w_1*$           & $w_3$          & $\boxed{w_2}*$   & $m_3$         & $m_1$         &$m_3$ &&& $w_2$ & \\

\end{tabular}
\end{center}
%\end{Example}

The matching labeled with boxes represents the matching $\varphi^{W}(\succ)$. 
%The table above displays the matching selected by the $W$-optimal mechanism according to the stated preferences $P=(P_{m_1}, P_{m_2}, P_{m_3},P_{w_1}, P_{w_2}, P_{w_3})$, marked in boxes, and denoted as $\varphi^{W}(P)$.
Meanwhile, the matching labeled with ``*" represents the matching $\varphi^{W}(P'_{m_2}, \succ_{-{m_2}})$.
However, this leads to:
$$\varphi_{m_2}^{W}(\succ)=w_1 \neq w_3= \varphi_{m_2}^{W}(P'_{m_2}, \succ_{-{m_2}}).$$
Then from the fact that $P'_{m_2}$ is a $\varphi^{W}$-boost misrepresentation of ${m_2}$ under the mechanism $\varphi^{W}$ when the true preference profile is $\succ$, we can conclude that $\varphi^{W}$ is not $\varphi^{W}$-boost-invariant.
 
\end{proof}

\begin{Remark}
Our Proposition \ref{prop:1} shows that: for any fixed mechanisim $\varphi$, our axiom $\varphi$-boost-invariance is strictly weaker than strategy-proofness; so our main theorem refines the impossibility result of \cite{roth1982economics}.
\end{Remark}

\begin{Remark}
The term truncation strategy has been used in two different senses --- one, by \cite{roth1991incentives} and the other by \cite{chen2017new}. 
Ours is the same as the latter. 
These two restrictions on strategy domain do not imply each other and $\varphi$-boost-invariance (which is defined based on $(\varphi,\succ)$-boost misrepresentation \`a la \cite{chen2017new}) is a strictly weaker axiom than strategy-proofness and truncation-invariance (which is defined based on truncation strategy \`a la \cite{roth1991incentives})(see section \ref{sec:2.1} for detailed discussions), 
which is why our negative result for one-to-one matching does not follow from the result for many-to-many matching in \cite{Jaramillo2013}, 
which uses the concept of \cite{roth1991incentives}.
\end{Remark}

%===============================================
\section{Conclusion}\label{sec:5}
%===============================================
%The relationship diagram in Figure \ref{fg:1} of section 1 summarizes the main results of this paper. 
As two important desiderata in one-to-one matching markets, stability and strategy-proofness are incompatible (\cite{roth1982economics}). The main focus of this paper lies on identifying the Machiavellian frontier of stable mechanisms with truthful revelation. We aim to explore the extent to which we can relax strategy-proofness while preserving  \cite{roth1982economics}'s impossibility theorem.

An agent may have an conjecture about which assignment would be assign to him/her by a matching mechanism. 
The axiom of $\varphi$-boost-invariance for a mechanism capture an intuition that a matchiing mechanism should incentive agents' willing to cooperate with the matchmaker while should not incentive misrepresentations.

Our Proposition \ref{prop:1} shows that the axiom of $\varphi$-boost-invariance is strictly weaker than the axiom of strategy-proofness.
%Specifically, we examine the potential benefits for agents in reporting a truncation of their true preferences at the point of truthful matching before the implementation of a stable mechanism in marriage markets introduced by \cite{gale1962college}). 
Our main result demonstrates that \cite{roth1982economics}'s impossibility theorem remains applicable even when we relax strategy-proofness to truncation-invariance. Since there exist some $\varphi$-boost-invariant mechanisms that violate strategy-proofness, such as the IA mechanism, our Theorem \ref{thm:impossibility} refines the Roth Impossibility Theorem.

Two potential directions for future research can be outlined as follows. 
Firstly, maintaining strategy-proofness, the concept of stability could be weakened in order to identify the fairness boundary that could potentially reverse Roth's impossibility theorem. 
This entails investigating the extent of fairness achievable under conditions necessitating truthful revelation. 
Secondly,  mechanisms could be made both Pareto-efficient (for oneside) and stable with respect to true preferences by relaxing strategy-proofness. This aims to address the paradox that Pareto-efficient matching (for students) may be unstable in the school choice problem. As \cite{abdulkadirouglu2003school} demonstrate, even the student-optimal stable matching may be Pareto dominated by unstable matchings. However, the incentives of the schools should not be considered in the school choice problem.
%This involves examining the extent to which a mechanism exists whose equilibrium outcome is Pareto-efficient for one side and stable with respect to true preferences, by restricting the strategy domain.
%\newpage

\bibliography{SCT}
\bibliographystyle{ecca}

\end{document}